\RequirePackage{fix-cm}
\documentclass[smallextended]{svjour3}       
\smartqed  
\usepackage{graphicx}
\usepackage{array}
\usepackage{amsmath}
\usepackage{lineno}
\nolinenumbers
\journalname{Journal}
\begin{document}

\title{On the verge of life: Distribution of nucleotide sequences in viral RNAs}


\author{Mykola Husev         \and
        Andrij Rovenchak 
}


\institute{M. Husev \email{mykola.husiev@lnu.edu.ua; mhusev@gmail.com} 
\and\\ A. Rovenchak \email{andrij.rovenchak@lnu.edu.ua; andrij.rovenchak@gmail.com} \at
Department for Theoretical Physics, Ivan Franko National University of Lviv\\ 
12 Drahomanov St, UA-79005, Lviv, Ukraine \\
}

\date{Received: date / Accepted: date}

\maketitle

\begin{abstract}
The aim of the study is to analyze viruses using parameters obtained from distributions of nucleotide sequences in the viral RNA. Seeking for the input data homogeneity, we analyze single-stranded RNA viruses only. Two approaches are used to obtain the nucleotide sequences; In the first one, chunks of equal length (four nucleotides) are considered. In the second approach, the whole RNA genome is divided into parts by adenine or the most frequent nucleotide as a ``space''. Rank--frequency distributions are studied in both cases. Within the first approach, the P\'olya and the negative hypergeometric distribution yield the best fit. For the distributions obtained within the second approach, we have calculated a set of parameters, including entropy, mean sequence length, and its dispersion. The calculated parameters became the basis for the classification of viruses. We observed that proximity of viruses on planes spanned on various pairs of parameters corresponds to related species. In certain cases, such a proximity is observed for unrelated species as well calling thus for the expansion of the set of parameters used in the classification. We also observed that the fourth most frequent nucleotide sequences obtained within the second approach are of different nature in case of human coronaviruses (different nucleotides for MERS, SARS-CoV, and SARS-CoV-2 versus identical nucleotides for four other coronaviruses). We expect that our findings will be useful as a supplementary tool in the classification of diseases caused by RNA viruses with respect to severity and contagiousness.
\end{abstract}

\section{Introduction}

Studies of genomes based on linguistic approaches date a few decades back [Brendel et al. 1986; Pevzner et al. 1989; Searls 1992; Botstein {\&} Cherry 1997; Gimona 2006; Falt\'ynek et al. 2019; Ji 2020]. An interplay with methods of statistical physics as well as theory of complex systems brought new insights into biology [Dehmer {\&} Emmert-Streib 2009; Qian 2013]. Studies range from attempted n-gram-based classification of genomes [Tomovi\'c et al. 2006; Huang {\&} Yu 2016] to algorithms for optimal segmentation of RNAs in secondary structure predictions [Licon et al. 2010] and analysis of substitution rates of coding genes during evolution [Lin et al. 2019], just to mention a few. Recently, neural networks and deep learning algorithms emerged as new tools to analyze nucleotide sequences [Fang et al. 2019; Singh et al. 2019; Merkus et al. 2020; Ren et al. 2020] offering wider prospects for studies of genomes. Viruses, balancing on the fuzzy border between non-alive and alive, hence remaining on the verge of life [Villarreal 2004; Kolb 2007; Carsetti 2020], are within the most interesting subjects of studies.

The aim of the present Letter is to draw attention to simple treatments of nucleotide sequences in viral RNAs by means of new parameters, which can be immediately extracted from genome data. We expect that such parameters can be potentially used as an auxiliary tool in the classification of viruses, cf, in particular, [Wang 2013]. The idea of this study is linked to the recent COVID-19 outbreak, and the analysis started from comparing human coronaviruses [Su et al. 2016; Wu et al. 2020] and some other viruses. To achieve relative homogeneity of the material, we restrict our sample to single-stranded RNA viruses only. Both positive- and negative-sense RNAs are considered. For future reference, we also include two retroviruses, HIV-1 and HIV-2.

The paper is organized as follows. Summary of data and description of methods are given in Section~\ref{sec2}. Results are presented in Section~\ref{sec3}. Finally, brief discussion is given in Section~\ref{sec4}. Detailed tables of numerical results are placed in the Appendix.

\section{Data and Methods}\label{sec2}

The viral genomes are taken from the databases of the National Center for Biotechnology Information (NCBI, https://www.ncbi.nlm.nih.gov); the complete list is given in Table~\ref{tab1}. Note that coronaviruses have rather long RNA genomes of ca. 30 kilobases (kba), which might bias the values of calculated parameters. To study the effect of RNA sizes, we also include some very short genomes, namely, Hepatitis D virus with 1682 ba [Saldanha et al. 1990] and Phage MS2 virus with 3569 ba [de Smit {\&} van Duin 1993], as well as two longest known RNA viruses, Ball python nidovirus with 33452 ba [Gorbalenya et al. 2006] and Planidovirus with 41178 ba [Saberi et al. 2018]. Still, the sizes of RNA viruses are much more homogeneous (the difference is up to 25 times) than those of DNA ones, which may vary by about four orders of magnitude [Campillo-Balderas et al. 2015].

\begin{table}[pt]
\caption{Viruses analyzed in the work. \label{tab1}}
{\scriptsize\begin{tabular}{r>{\raggedright}p{2cm}>{\raggedright}p{4.5cm}crll} 
\hline\noalign{\smallskip}
No.	&	Short name	&	Full name	&	Type$^{a}$	&	\parbox[t]{1cm}{\ \ Size \\ (bases)}	&	NCBI source$^{b}$	\\
\noalign{\smallskip}\hline\noalign{\smallskip}
1	&	A/H1N1	&	Influenza A virus (A/swine/La~Habana/ 130/2010(H1N1))	&	$(-)$	&	13371	&	\parbox[t]{2cm}{HE584753.1\\\ldots \\ HE584760.1}	\\[24pt]
2	&	Ball python nidovirus	&	Ball python nidovirus 1	&	(+)	&	33452	&	674660326	\\[12pt]
3	&	Dengue	&	Dengue virus 2	&	(+)	&	10723	&	158976983	\\
4	&	Ebola	&	Zaire ebolavirus	&	$(-)$	&	18962	&	MK672824.1	\\
5	&	Feline-CoV	&	Feline infectious peritonitis virus	&	(+)	&	29355	&	315192962	\\
6	&	HCoV-229E	&	Human coronavirus 229E	&	(+)	&	27317	&	12175745	\\
7	&	HCoV-HKU1	&	Human coronavirus HKU1	&	(+)	&	29926	&	85667876	\\
8	&	HCoV-NL63	&	Human coronavirus NL63	&	(+)	&	27553	&	49169782	\\
9	&	HCoV-OC43	&	Human coronavirus OC43	&	(+)	&	30741	&	1578871709	\\
10	&	Hepatitis A	&	Hepatovirus A	&	(+)	&	7478	&	NC\_001489.1	\\
11	&	Hepatitis C	&	Hepatitis C virus genotype 1	&	(+)	&	9646	&	22129792	\\
12	&	Hepatitis D	&	Hepatitis delta virus	&	$(-)$	&	1682	&	13277517	\\
13	&	Hepatitis E	&	Hepatitis E virus	&	(+)	&	7176	&	NC\_001434.1	\\
14	&	HIV-1	&	Human immunodeficiency virus 1	&	(retro)	&	9181	&	9629357	\\
15	&	HIV-2	&	Human immunodeficiency virus 2	&	(retro)	&	10359	&	9628880	\\
16	&	HRV-A	&	Human rhinovirus A1	&	(+)	&	7137	&	1464306962	\\
17	&	HRV-B	&	Human rhinovirus B3	&	(+)	&	7208	&	1464306975	\\
18	&	HRV-C	&	Human rhinovirus NAT001	&	(+)	&	6944	&	1464310212	\\
19	&	Marburg	&	Lake Victoria marburgvirus - Ravn	&	$(-)$	&	19114	&	DQ447649.1	\\
20	&	Measles	&	Measles virus strain Edmonston	&	$(-)$	&	15894	&	AF266290.1	\\
21	&	MERS	&	Middle East respiratory syndrome coronavirus	&	(+)	&	30119	&	667489388	\\
22	&	Norovirus	&	Norovirus Hu/GI.1/ CHA6A003\_20091104/2009/USA	&	(+)	&	7600	&	KF039737.1	\\
23	&	Phage MS2	&	Enterobacteria phage MS2	&	(+)	&	3569	&	176120924	\\
24	&	Planidovirus	&	Planarian secretory cell nidovirus	&	(+)	&	41178	&	1571803928	\\
25	&	Polio	&	Poliovirus (Enterovirus C)	&	(+)	&	7440	&	NC\_002058.3	\\
26	&	Rabies	&	Rabies virus strain SRV9	&	$(-)$	&	11928	&	AF499686.2	\\
27	&	SARS	&	Severe acute respiratory syndrome coronavirus	&	(+)	&	29751	&	30271926	\\
28	&	SARS-CoV-2	&	Severe acute respiratory syndrome coronavirus 2	&	(+)	&	29903	&	NC\_045512	\\
29	&	Yellow fever	&	Yellow fever virus	&	(+)	&	10862	&	NC\_002031.1 g-max	\\
30	&	Zika	&	Zika virus	&	(+)	&	10794	&	226377833 g-max	\\
\noalign{\smallskip}\hline\noalign{\smallskip}
\end{tabular} }
Notes:\\
$^{a}$ Negative-sense RNA $(-)$, positive-sense RNA (+) or retro.\\
$^{b}$ All addresses should be prefixed by https://www.ncbi.nlm.nih.gov/nuccore/.
\end{table}

We use two approaches to define nucleotide sequences. The first one is based on cutting an RNA genome into chunks of equal length of $n$ nucleotides. The second approach is rooted in linguistics, so that the most frequent nucleotide is treated as a ``space'' dividing a RNA into ``words'' of different lengths [Rovenchak 2018]. Note also distantly related units applied in the analysis of the human DNA, so called motifs [Liang 2014].

To demonstrate the first approach, with equal-length chunks, let us consider the Ebolavirus genome, starting with the following nucleotide sequence:

\begin{equation}\label{eq1}
\textrm{GGACACACAAAAAGAAAGAAGAATTTTTAGGATCTTTTGT\ldots}\ .
\end{equation} 
Choosing the chunk length $n = 4$, we obtain:
\begin{equation}\label{eq2}
\textrm{GGAC\ ACAC\ AAAA\ AGAA\ AGAA\ GAAT\ TTTT\ AGGA\ TCTT\ TTGT\ \ldots}\ .
\end{equation}
Eventually, for RNA length not being multiples of four, the last chunk can have one to three nucleotides. Obviously, the number of all possible 4-nucleotide combinations is $4^4 = 256$. Note that longer chunks would yield much higher variety of combinations with frequencies being distributed very smoothly. On the other hand, we would like to avoid studies of shorter chunks, like three-nucleotide sequences corresponding to codons. So, the length $n = 4$ seems optimal for our analysis.

In the second approach, the same Ebolavirus sequence (]ref{eq1}) can be split using the most frequent nucleotide -- adenine -- as a ``space'' into the following:
\begin{equation}\label{eq3}
\textrm{GG C C C X X X X G X X G X G X TTTTT GG TCTTTTGT\ldots}\ .
\end{equation}
The ``X'' stands for a zero-length element inserted between two consecutive ``A''s.

We have also applied peculiar treatment of the Influenza A virus (H1N1) by adding spaces between each of eight segments of its RNA in the first and second approaches.

In both approaches, we calculate the frequencies of obtained nucleotide chunks within a given genome split in the respective manner and compile the rank--frequency distributions. The latter are obtained in a standard manner as follows: the most frequent item has rank 1, the second most frequent one has rank 2 and so on. Items with equal frequencies are given consecutive ranks in a random order, which is not relevant.

\section{Results}\label{sec3}

The rank--frequency distributions obtained using the first approach -- with 4-nucleotide chunks -- were analyzed using a special software, AltmannFitter 2.1 [Altmann 2000]. We found that two discrete distributions describe the obtained data with the highest precision, so called 1-displaced negative hypergeometric distribution [Grzybek 2007; Wilson 2013]:
\begin{equation}\label{eq4}
p_r=\frac{ \binom{M+r-2}{r-1} \binom{K-M+n-r-2}{n-r+1} }{ \binom{K+n-1}{n} },
\qquad r=1,2,3,\ldots\ .
\end{equation}
and P\'olya distribution [Wimmer {\&} Altmann 1999; Johnson et al. 2005]:
\begin{equation}\label{eq5}
p_r=\frac{ \binom{-p/s}{r-1} \binom{(p-1)/s}{n-r+1} }{ \binom{-1/s}{n} },
\qquad r=1,2,3,\ldots\ .
\end{equation}
Absolute frequencies are obtained by multiplying $p_r$ by the sample size $N$. In most cases, the discrepancy coefficient $C = \chi^2/N$ is smaller than 0.02, which is considered a good fit [Ma\v{c}utek 2008]. Typical rank--frequency distributions and respective fits are shown in Figure~\ref{fig1}. Complete data are summarized in Table~\ref{tab2} in the Appendix and visualized in Figure~\ref{fig2}.

\begin{figure}[th]
\centerline{\includegraphics[width=6cm]{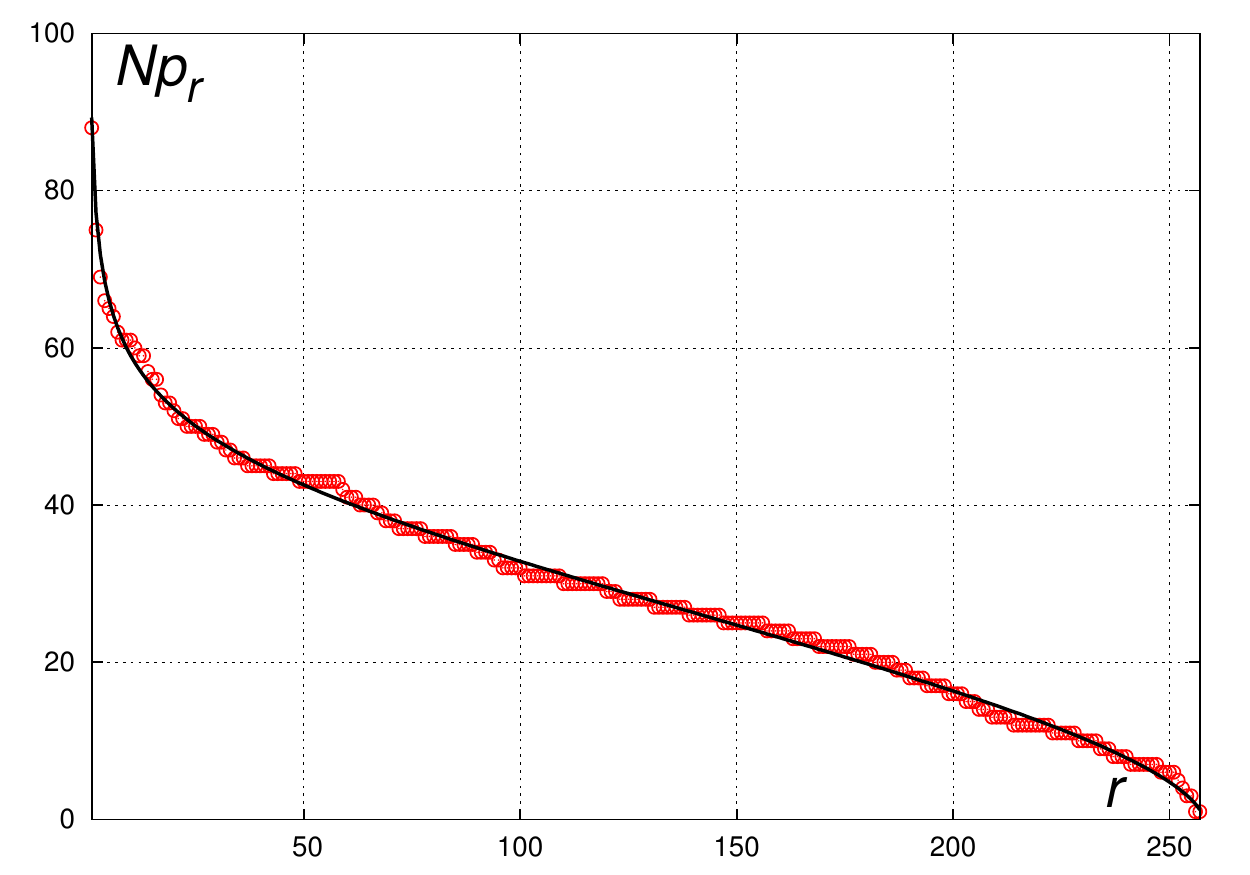}\quad \includegraphics[width=6cm]{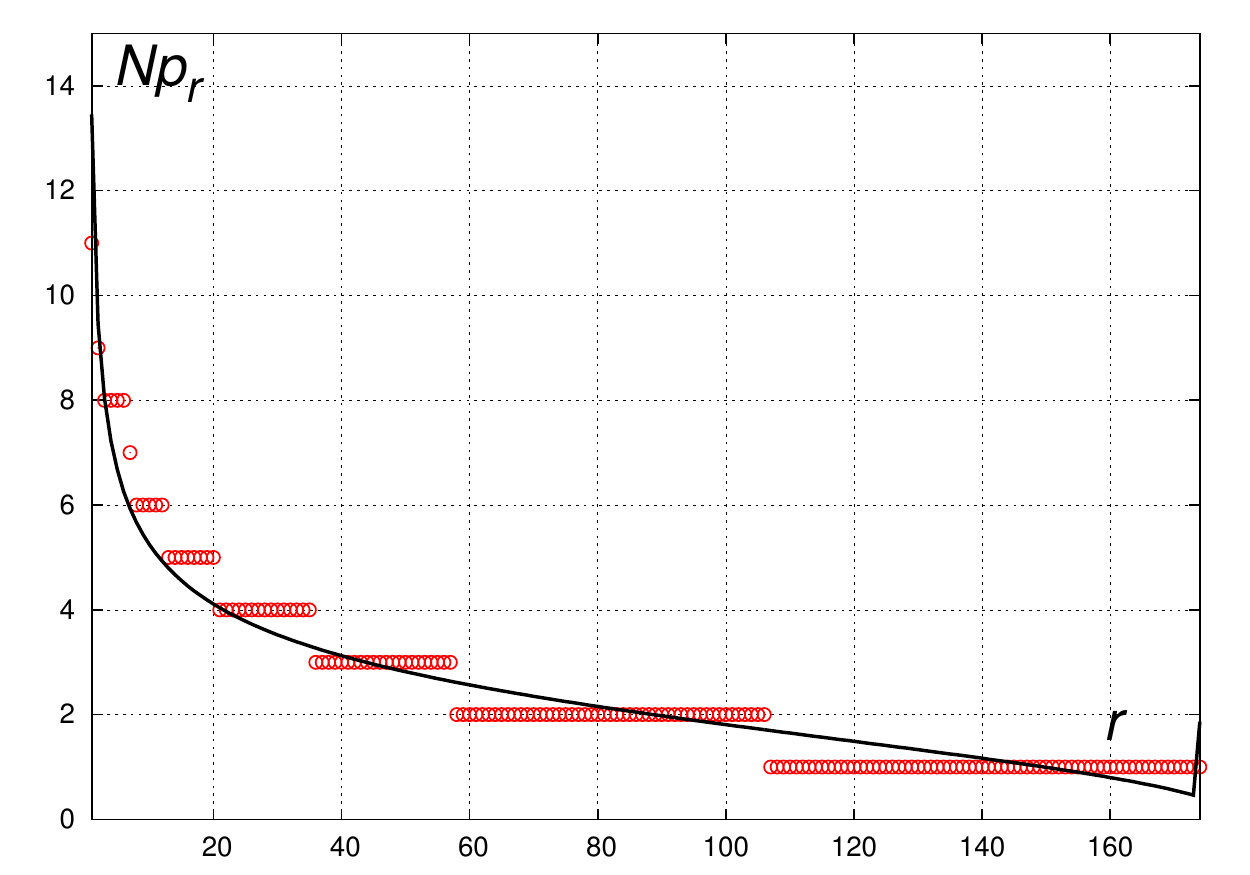}}
\caption{Typical rank--frequency distributions and respective fits. The left panel shows the data for MERS and the fit with the hypergometric distribution, which is one of the best ($C = 0.0011$). The right panel demonstrated the worst fit obtained for the Hepatitis D virus data fit with the P\'olya distribution ($C = 0.0342$).\label{fig1}}
\end{figure}

\begin{figure}[th]
\centerline{\includegraphics[width=6cm]{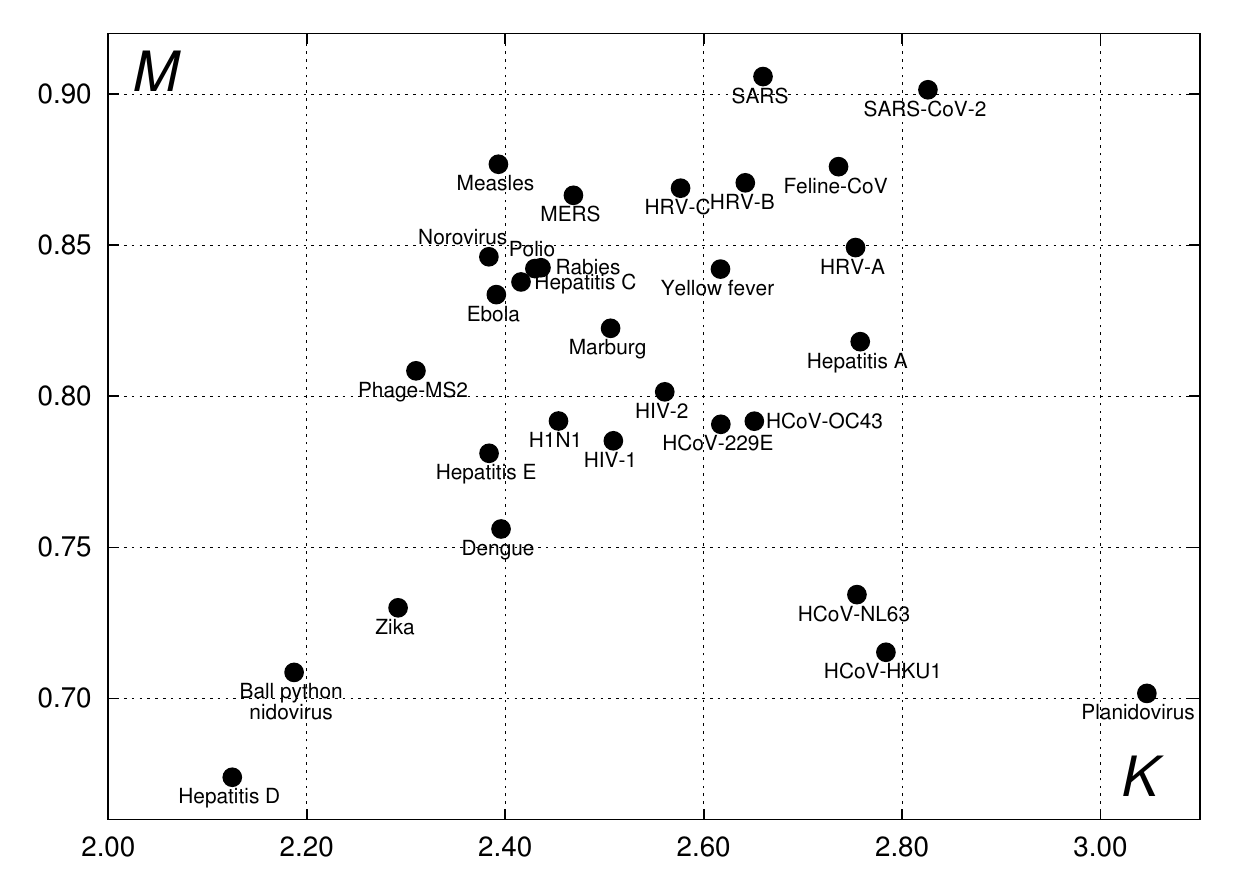}\quad \includegraphics[width=6cm]{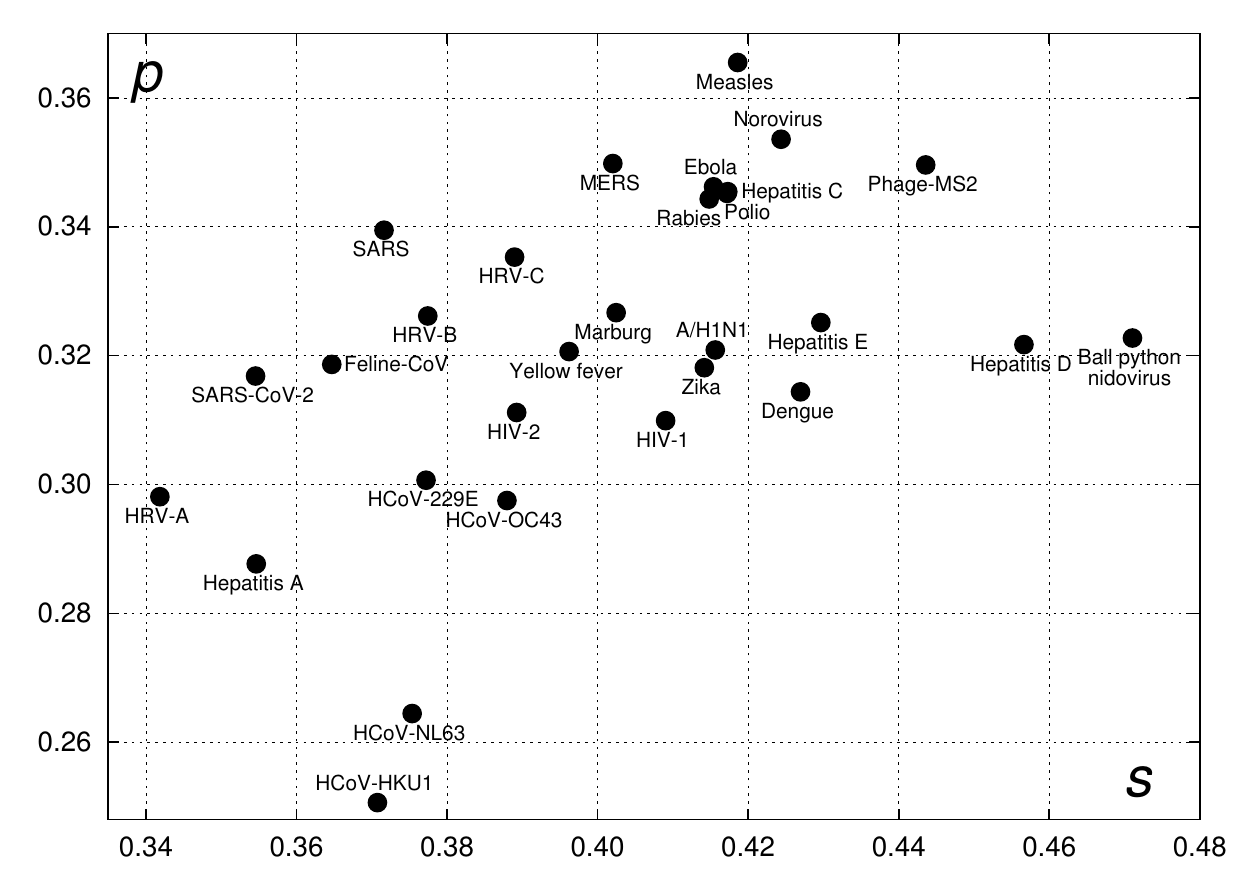}}
\caption{Location of viruses on the $K - M$ plane (negative hypergeometric fit, left panel) 
and $s - p$ plane (P\'olya fit, right panel).\label{fig2}}
\end{figure}

The first immediate observation from Figure~\ref{fig2} is that the length of genomes has no special influence on the fitting parameters. Indeed, both the shortest Hepatitis D genome and two longest -- Ball python nidovirus and Planidovirus --  genomes have close values of $M$ or $s$ parameters. On  the other hand, for genomes of similar lengths (coronaviruses) a clear separation is seen with respect to $M$ and $p$ parameters. It is even more pronounced in the former case corresponding to the negative hypergeometric distribution: lower values for HCoV viruses (229E, HKU1, NL63, and OC43) and higher ones for MERS, SARS, and SARS-CoV-2.

\begin{figure}[th]
\centerline{\includegraphics[width=8cm]{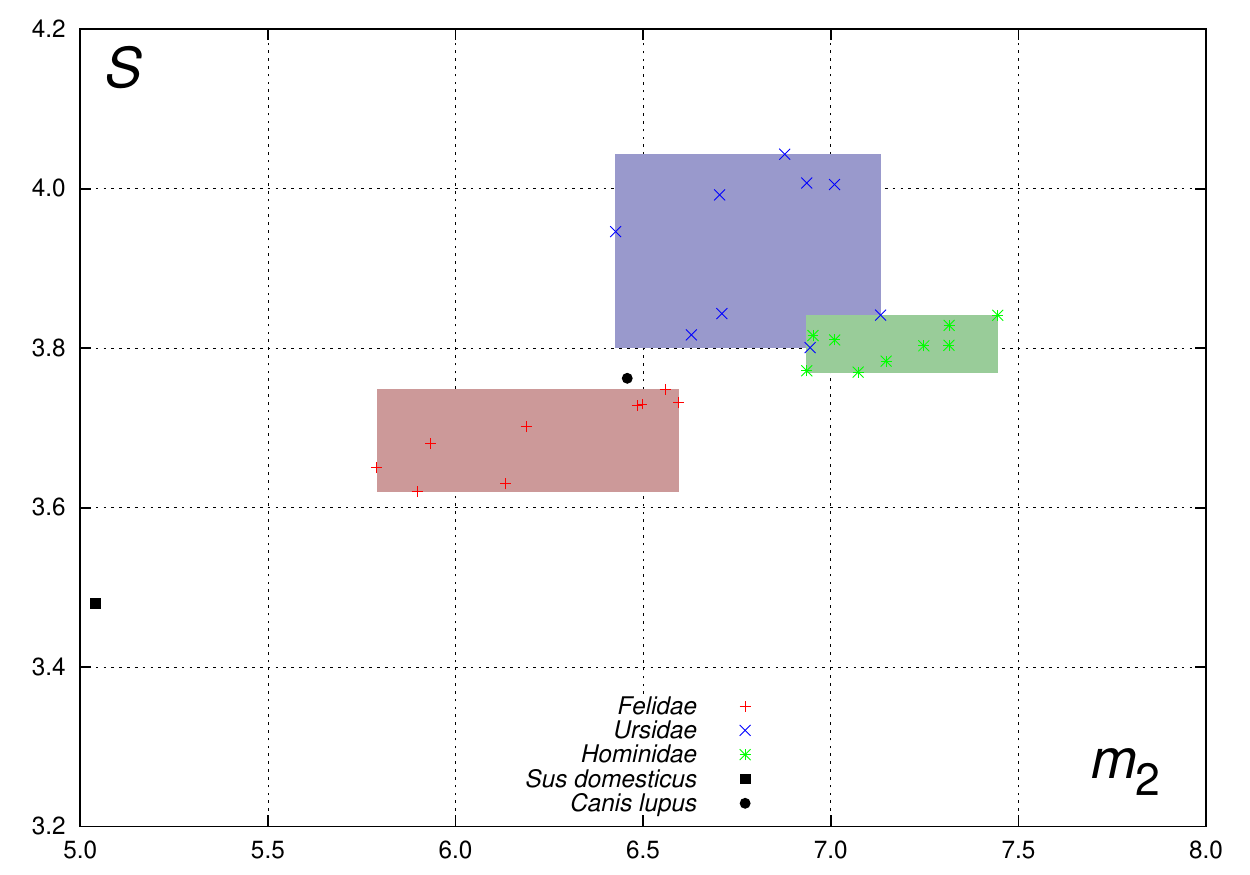}}
\caption{Grouping of mammal species on the $m_2 - S$ plane. Red-shaded area corresponds to \textit{Felidae}, the blue one denotes \textit{Ursidae}, and the green-one corresponds to \textit{Hominidae}. Calculations are made using mitochondrial DNAs with adenine as a ``space''.\label{fig3}}
\end{figure}

Rank--frequency distributions were also compiled for nucleotide ``words'' obtained using the second approach and used to calculate certain parameters, like entropy, mean length (first central moment), length dispersion (second central moment) and some others. Previous studies [Rovenchak 2018] showed that entropy and mean lengths of nucleotide sequences in the mitochondrial DNA can be used to distinguish species and genera of mammals. It appears, however, that even better results are achieved with the ``entropy -- length dispersion'' pair of variables, cf. Figure~\ref{fig3}.

The parameters are defined as follows. Entropy is given by
\begin{equation}\label{eq6}
S=-\sum_{r=1}^{r_{\rm max}} p_r \ln p_r ,
\end{equation}
where the upper summation limit corresponds to the total number of different ``words'' in the list and relative frequencies $p_r$ are
\begin{equation}\label{eq7}
p_r=f_r/N,\qquad\textrm{where} N=\sum_r f_r
\end{equation}
and $f_r$ are absolute frequencies at rank $r$. Mean length and length dispersion are
\begin{equation}\label{eq8}
m_1=\frac{1}{N} \sum_i x_i, \qquad
m_2=\frac{1}{N} \sum_i (x_i-m_1)^2.
\end{equation}
where the summations run over all the ``words'' of the analyzed genome. Lengths $x_i$ of a particular word are counted as the number of nucleotides except for ``X'' having length zero.

One should note that from similarity of species one can expect proximity of points but not vice versa: it would be too bold to expect species distinguishability from only two parameters.

\begin{figure}[th]
\centerline{\includegraphics[width=8cm]{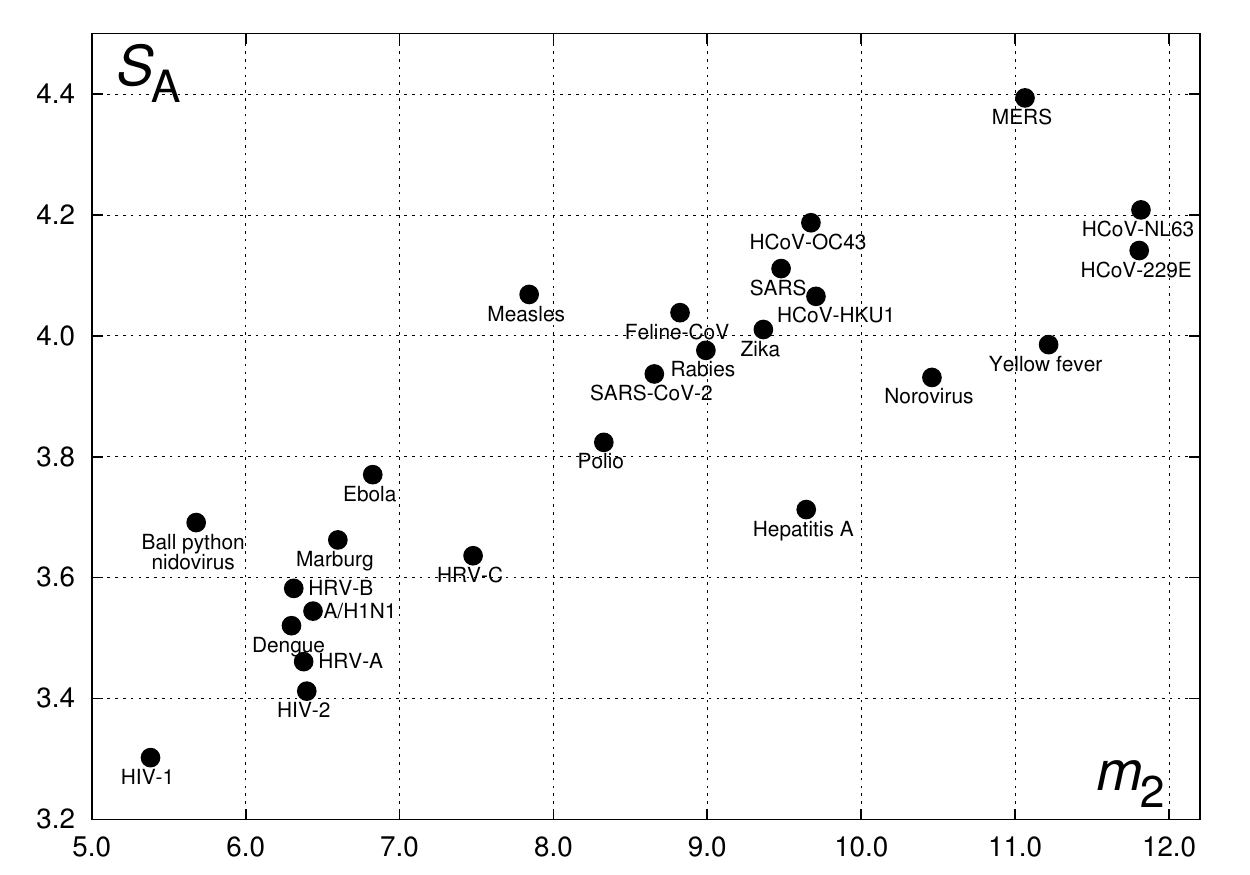}}
\caption{Location of viruses on the $m_2 - S$ plane. Calculations are made using RNAs with adenine as a ``space'', hence entropy is denoted $S_{\rm A}$.\label{fig4}}
\end{figure}

\begin{figure}[th]
\centerline{\includegraphics[width=8cm]{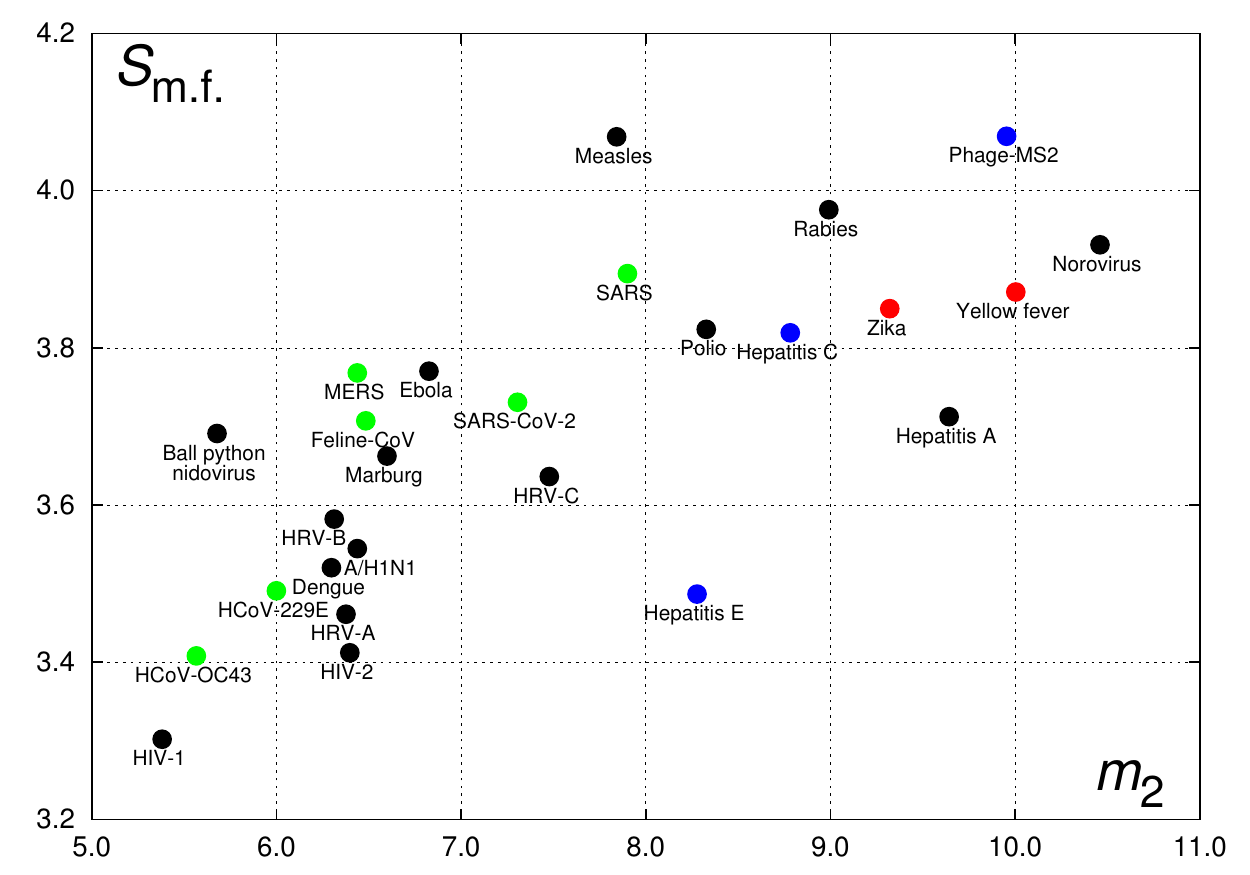}}
\caption{Location of viruses on the $m_2 - S$ plane. Calculations are made using RNAs with the most frequent nucleotide as a ``space'', hence entropy is denoted $S_{\rm m.f.}$.\label{fig5}}
\end{figure}

\begin{figure}[th]
\centerline{\includegraphics[width=8cm]{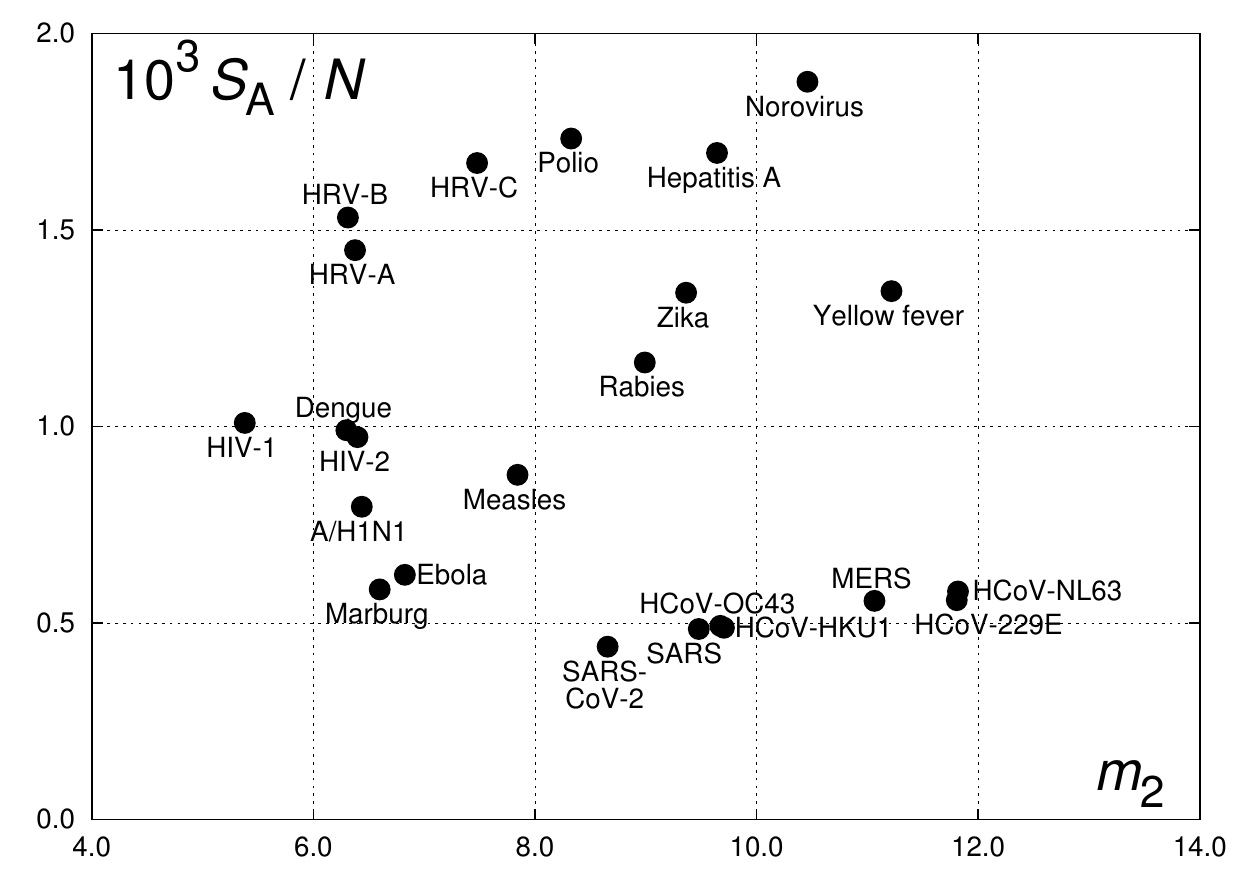}}
\caption{Location of viruses on the $m_2 - S/N$ plane. Calculations are made using RNAs with adenine as a ``space'', hence entropy is denoted $S_{\rm A}$. The vertical axis thus represents the entropy divided by the number of nucleotide sequences separated by adenine in the respective genome.\label{fig6}}
\end{figure}

This second approach can be divided into two sub-branches: (a) adenine, which is the most frequent nucleotide in most species studied in the present work, is used as a ``space''; (b) the most frequent nucleotide is used as a ``space''. The latter is mostly relevant for RNAs, where low frequencies of adenine yield too long ``words'' thus significantly distorting the expected dependencies. The respective results are shown in Figures~\ref{fig4}--\ref{fig6}. All the data are summarized in Table~\ref{tab3}.

In Figure~\ref{fig6}, we can observe in particular that $\alpha$-coronaviruses, HCoV-229E and HCoV-NL63, have very close values of the parameters (the respective point nearly overlap). A similar situation is with $\beta$-corovaniruses HCoV-OC43 and HCoV-HKU1. Two other $\beta$-corovaniruses, SARS and SARS-CoV-2, are located close to HCoV-OC43 and HCoV-HKU1, while MERS occupies an intermediate position. The latter virus also significantly differs in the entropy value, see Figure~\ref{fig5}. On the other hand, calculations with the most frequent nucleotide used as a space (T for the analyzed coronaviruses) do not exhibit such a grouping, see Figure~\ref{fig4}.

When looking in detail into the rank--frequency distributions corresponding to coronaviruses we have discovered the following pattern: the first rank is always occupied by ``X'' followed by three single-nucleotide ``words'' with ranks 2--4, while the fifth ranks are occupied by a two-nucleotide sequence with either the same (4-same) or different (4-diff) nucleotides, see Table~\ref{tab4}. Curiously, different nucleotides correspond to coronaviruses causing much more severe diseases. This observation is yet to be extended onto a wider material, but the preliminary data for the analyzed human viruses are as follows:
\begin{itemize}
\item \textbf{4-same:} Dengue, HCoV-229E, HCoV-HKU1, HCoV-NL63, HCoV-OC43, HIV-1, HIV-2, HRV-A, HRV-B, HRV-C, Polio;
\item \textbf{4-diff:} A/H1N1, Ebola, Hepatitis A, Hepatitis C, Hepatitis E, Marburg, Measles, MERS, Norovirus, Rabies, SARS, SARS-CoV-2.
\end{itemize}
Three other viruses, Hepatitis D, Yellow fever, and Zika, do not follow either pattern having a two-nucleotide sequence with as low ranks as 3 or 4.

\begin{table}[pt]
\caption{Top-ranked nucleotide sequences in the genomes of the human coronaviruses. \label{tab4}}
{\scriptsize\begin{tabular}{r||lr|lr|lr|lr|lr|lr|lr}
\hline\noalign{\smallskip}
		&	\multicolumn{2}{c|}{MERS} & \multicolumn{2}{c|}{SARS} & \multicolumn{2}{c|}{SARS-CoV-2} & \multicolumn{2}{c|}{HCoV-229E} & \multicolumn{2}{c|}{HCoV-HKU1} & \multicolumn{2}{c|}{HCoV-NL63} & \multicolumn{2}{c}{HCoV-OC43}	\\
$r$	&	``word''	&	 $f_r$	&	``word''	&	 $f_r$	&	``word''	&	$f_r$	&	``word''	&	$f_r$	&	``word''	&	$f_r$	&	``word''	&	 $f_r$	&	``word''	&	$f_r$	\\
\noalign{\smallskip}\hline\noalign{\smallskip}
1	&	X	&	3098	&	X	&	2845	&	X	&	3215	&	X	&	3380	&	X	&	4694	&	X	&	4272	&	X	&	3895	\\
2	&	G	&	876	&	G	&	795	&	G	&	858	&	G	&	1033	&	A	&	1183	&	G	&	1149	&	G	&	1105	\\
3	&	A	&	701	&	C	&	568	&	A	&	623	&	A	&	615	&	G	&	1151	&	A	&	814	&	A	&	963	\\
4	&	C	&	668	&	A	&	567	&	C	&	542	&	C	&	458	&	C	&	581	&	C	&	521	&	C	&	468	\\
5	&	GC	&	256	&	GC	&	316	&	GC	&	255	&	GG	&	288	&	AA	&	399	&	GG	&	387	&	AA	&	324	\\
6	&	GG	&	234	&	GA	&	217	&	GG	&	245	&	GC	&	284	&	GA	&	339	&	AA	&	318	&	GA	&	322	\\
7	&	GA	&	223	&	GG	&	202	&	AA	&	218	&	AA	&	211	&	GG	&	338	&	GA	&	296	&	GG	&	293	\\
8	&	AA	&	214	&	AC	&	196	&	AC	&	214	&	GA	&	210	&	AC	&	271	&	GC	&	232	&	GC	&	269	\\
9	&	AC	&	194	&	AA	&	167	&	GA	&	208	&	AC	&	156	&	AG	&	227	&	AG	&	194	&	AC	&	190	\\
10	&	AG	&	134	&	CA	&	154	&	AG	&	138	&	AG	&	128	&	GC	&	223	&	AC	&	190	&	AG	&	171	\\
11	&	CC	&	131	&	AG	&	102	&	CA	&	127	&	CA	&	105	&	AAA	&	117	&	CA	&	107	&	CA	&	96	\\
12	&	CA	&	126	&	CC	&	81	&	CC	&	79	&	CC	&	56	&	CA	&	113	&	CC	&	69	&	CC	&	86	\\
13	&	CG	&	80	&	CG	&	74	&	AAA	&	64	&	GAC	&	52	&	CC	&	104	&	CG	&	58	&	AAA	&	76	\\
\noalign{\smallskip}\hline
\end{tabular} }
\end{table}

\section{Discussion}\label{sec4}

We have presented several possible approaches to simple parametrization of RNA viruses based on the analysis of nucleotide sequences in viral genomes. They are based on discrete distributions (negative hypergeometric and P\'olya) for equal-length (4-nucleotide) chunks and on the pair ``entropy -- length dispersion'' for distributions of sequences separated by adenine or another most frequent nucleotide. Related viruses are characterized by close values of the calculated parameters. In some cases, similar values are also obtained for unrelated viruses. This is not surprising as representing viruses on a plane means a two-parametric projection of points that are certainly described by more than two variables. We consider our study as preliminary steps in discovering such variables.

Observations regarding peculiarities of rank--frequency distributions, with the fourth most frequent sequence containing two either the same or different nucleotides (4-same vs 4-diff), support the fact that 4-diff cases correspond to viruses causing potentially more severe diseases when dealing with seven human coronaviruses. This tendency is generally preserved if the analyzed set is expanded by other viruses studied in this work. Some precautions concern, in particular, the two HIV types, which fall into the 4-same category while certainly being extremely dangerous. However, HIV are not strictly RNA viruses but retroviruses, so we suggest that the reported peculiarities might be specific for RNA viruses only. ``False-positive'' alerts (cf. Norovirus in the 4-diff category) are not problematic, but the rate of ``false-negative'' results (severe diseases in the 4-same category) is yet to be identified. Expansion of the analyzed material in future studies would help to clarify the relevance of this observation. To establish relations between peculiarities of the rank--frequency distributions in virus genomes and disease severity, a formalization of the latter is required. Initially we planned using the case fatality rate (CFR) indicator [Reich et al. 2012; Kim et al. 2020] but where not able to find a study with data for different viruses based on a unified approach, similar, e.g., to [GBD 2017].

The main expected outcome of our reported analysis is a call for collaboration to expand the dataset and consistently classify diseases caused by RNA viruses, in particular with respect to severity and contagiousness. If some simple patterns could be established in the nucleotide distributions, this might help alerting healthcare systems, which seems to become a very topical issue from this year on.

\section*{Conflict of interest}

The authors, Mykola Husev and Andrij Rovenchak, declare that they have no conflict of interest.

\section*{Ethical approval}

This article does not contain any studies with human participants or animals performed by any of the authors.

\clearpage
\appendix

\section{Tables of data}

\begin{table}[h]
\caption{Fitting parameters for the distributions of four-nucleotide chunks. \label{tab2}}
{\begin{tabular}{l|c|c|cccc|cccc}
\hline\noalign{\smallskip}
Virus	&	Entropy  &	Size	& \multicolumn{4}{c|}{Negative hypergeometric distribution}	&						 \multicolumn{4}{c}{P\'olya distribution}\\
\cline{4-7}\cline{8-11}\noalign{\smallskip}
	& $S_4$	&  (chunks) &	$K$	&	$M$	&	$n$	&	$C$	&	$s$	&	$p$	&	$n$	&	$C$	\\
\noalign{\smallskip}\hline\noalign{\smallskip}
A/H1N1	&	5.3515	&	3345	&	2.4536	&	0.7918	&	258	&	0.0057	&	0.4156	&	0.3209	&	259	&	0.006	\\
Ball python nidov.	&	5.3385	&	8363	&	2.1873	&	0.7087	&	256	&	0.0043	&	0.4711	&	0.3227	&	256	&	0.005	\\
Dengue	&	5.3219	&	2681	&	2.3958	&	0.7561	&	256	&	0.0051	&	0.427	&	0.3144	&	256	&	0.0055	\\
Ebola	&	5.4002	&	4741	&	2.3911	&	0.8336	&	256	&	0.0008	&	0.4154	&	0.3462	&	256	&	0.001	\\
Feline-CoV	&	5.3357	&	7339	&	2.7359	&	0.8759	&	257	&	0.0011	&	0.3647	&	0.3186	&	257	&	0.0011	\\
HCoV-229E	&	5.2899	&	6830	&	2.6172	&	0.7907	&	256	&	0.0014	&	0.3772	&	0.3007	&	257	&	0.0013	\\
HCoV-HKU1	&	5.1491	&	7482	&	2.7836	&	0.7153	&	260	&	0.0057	&	0.3707	&	0.2506	&	261	&	0.007	\\
HCoV-NL63	&	5.1738	&	6889	&	2.7545	&	0.7344	&	256	&	0.0035	&	0.3754	&	0.2644	&	255	&	0.0042	\\
HCoV-OC43	&	5.2854	&	7686	&	2.651	&	0.7918	&	258	&	0.0027	&	0.3879	&	0.2975	&	257	&	0.0031	\\
Hepatitis A	&	5.1923	&	1870	&	2.7578	&	0.818	&	239	&	0.0079	&	0.3546	&	0.2877	&	243	&	0.0075	\\
Hepatitis C	&	5.3871	&	2412	&	2.4158	&	0.8378	&	254	&	0.0029	&	0.4173	&	0.3454	&	254	&	0.003	\\
Hepatitis D	&	4.9309	&	421	&	2.1249	&	0.6739	&	178	&	0.0333	&	0.4566	&	0.3217	&	178	&	0.0342	\\
Hepatitis E	&	5.3405	&	1794	&	2.3837	&	0.7811	&	254	&	0.007	&	0.4297	&	0.3251	&	254	&	0.0077	\\
HIV-1	&	5.2425	&	2296	&	2.509	&	0.7853	&	239	&	0.006	&	0.409	&	0.3099	&	239	&	0.0066	\\
HIV-2	&	5.3114	&	2590	&	2.5607	&	0.8015	&	256	&	0.003	&	0.3892	&	0.3112	&	256	&	0.0029	\\
HRV-A	&	5.2618	&	1785	&	2.753	&	0.8492	&	248	&	0.0081	&	0.3418	&	0.2981	&	254	&	0.0077	\\
HRV-B	&	5.2793	&	1802	&	2.6419	&	0.8706	&	238	&	0.0043	&	0.3774	&	0.3262	&	239	&	0.0042	\\
HRV-C	&	5.3165	&	1736	&	2.5766	&	0.8688	&	243	&	0.0033	&	0.389	&	0.3353	&	243	&	0.0033	\\
Marburg	&	5.3418	&	4779	&	2.5061	&	0.8225	&	252	&	0.002	&	0.4025	&	0.3267	&	253	&	0.0021	\\
Measles	&	5.4293	&	3974	&	2.3932	&	0.8767	&	256	&	0.0022	&	0.4186	&	0.3655	&	256	&	0.0022	\\
MERS	&	5.4040	&	7530	&	2.4687	&	0.8665	&	256	&	0.0011	&	0.402	&	0.3498	&	257	&	0.0013	\\
Norovirus	&	5.4015	&	1900	&	2.3835	&	0.8461	&	253	&	0.0046	&	0.4244	&	0.3536	&	254	&	0.0045	\\
Phage-MS2	&	5.3680	&	893	&	2.31	&	0.8084	&	249	&	0.0107	&	0.4436	&	0.3496	&	249	&	0.0114	\\
Planidovirus	&	5.0360	&	10295	&	3.0466	&	0.7017	&	261	&	0.0183	&	0.2449	&	0.1863	&	315	&	0.0161	\\
Polio	&	5.3837	&	1860	&	2.43	&	0.8423	&	254	&	0.0044	&	0.4172	&	0.3452	&	254	&	0.0047	\\
Rabies	&	5.3802	&	2982	&	2.4359	&	0.8425	&	252	&	0.0027	&	0.4148	&	0.3443	&	253	&	0.0028	\\
SARS	&	5.3825	&	7438	&	2.6599	&	0.9058	&	256	&	0.002	&	0.3716	&	0.3395	&	257	&	0.0019	\\
SARS-CoV-2	&	5.3330	&	7476	&	2.826	&	0.9014	&	258	&	0.0022	&	0.3546	&	0.3168	&	258	&	0.0021	\\
Yellow fever	&	5.3430	&	2716	&	2.6168	&	0.8421	&	258	&	0.005	&	0.3962	&	0.3206	&	257	&	0.0059	\\
Zika	&	5.3377	&	2699	&	2.2919	&	0.73	&	257	&	0.0081	&	0.4142	&	0.3181	&	258	&	0.0053	\\
\noalign{\smallskip}\hline\noalign{\smallskip}
\end{tabular} }
Note: 
Entropies $S_4$ are calculated for the distributions of four-nucleotide chunks using Equation~(\ref{eq6}).
\end{table}

\begin{table}[h]
\caption{Parameters for the distributions of nucleotide sequences separated by a specific nucleotide	\label{tab3}}
{\begin{tabular}{lcrrcc} 
\hline\noalign{\smallskip}
Virus&	Entropy	&Size\ \ \ \ \ &	Size\ \ \ \ &	Mean length &	Length dispersion \\
	&	$S$	&	(``words'')	&	(bases)	& $m_1$ & $m_2$ \\
\noalign{\smallskip}\hline\noalign{\smallskip}
\multicolumn{6}{l}{\textbf{A considered a ``space'' even if not being the most frequent:}}\\
\noalign{\smallskip}\hline\noalign{\smallskip}
A/H1N1	&	3.5446	&	4456	&	13371	&	2.0025	&	6.4378	\\
Ball python nidovirus	&	3.6911	&	11118	&	33452	&	2.0089	&	5.6785	\\
Dengue	&	3.5204	&	3554	&	10723	&	2.0174	&	6.2980	\\
Ebola	&	3.7703	&	6056	&	18962	&	2.1313	&	6.8261	\\
Feline-CoV	&	4.0381	&	8572	&	29355	&	2.4246	&	8.8222	\\
HCoV-229E	&	4.1411	&	7421	&	27317	&	2.6812	&	11.8059	\\
HCoV-HKU1	&	4.0653	&	8332	&	29926	&	2.5918	&	9.7058	\\
HCoV-NL63	&	4.2082	&	7254	&	27553	&	2.7985	&	11.8171	\\
HCoV-OC43	&	4.1871	&	8503	&	30741	&	2.6154	&	9.6729	\\
Hepatitis A	&	3.7125	&	2189	&	7478	&	2.4166	&	9.6428	\\
Hepatitis C	&	4.7418	&	1890	&	9646	&	4.0349	&	23.7224	\\
Hepatitis D	&	3.7600	&	340	&	1682	&	3.9500	&	30.1122	\\
Hepatitis E	&	4.9569	&	1231	&	7176	&	4.8302	&	30.5829	\\
HIV-1	&	3.3022	&	3273	&	9181	&	1.8054	&	5.3819	\\
HIV-2	&	3.4121	&	3507	&	10359	&	1.9541	&	6.3979	\\
HRV-A	&	3.4610	&	2389	&	7137	&	1.9879	&	6.3770	\\
HRV-B	&	3.5822	&	2339	&	7208	&	2.0821	&	6.3131	\\
HRV-C	&	3.6362	&	2177	&	6944	&	2.1902	&	7.4778	\\
Marburg	&	3.6623	&	6256	&	19114	&	2.0555	&	6.5991	\\
Measles	&	4.0685	&	4639	&	15894	&	2.4264	&	7.8423	\\
MERS	&	4.3936	&	7901	&	30119	&	2.8122	&	11.0646	\\
Norovirus	&	3.9312	&	2094	&	7600	&	2.6299	&	10.4595	\\
Phage MS2	&	4.1385	&	836	&	3569	&	3.2703	&	15.0130	\\
Planidovirus	&	3.0356	&	16361	&	41178	&	1.5169	&	3.6413	\\
Polio	&	3.8237	&	2207	&	7440	&	2.3715	&	8.3277	\\
Rabies	&	3.9758	&	3419	&	11928	&	2.4890	&	8.9910	\\
SARS	&	4.1112	&	8482	&	29751	&	2.5077	&	9.4794	\\
SARS-CoV-2	&	3.9369	&	8955	&	29903	&	2.3394	&	8.6559	\\
Yellow fever	&	3.9853	&	2964	&	10862	&	2.6650	&	11.2174	\\
Zika	&	4.0105	&	2992	&	10794	&	2.6080	&	9.3647	\\
\noalign{\smallskip}\hline\noalign{\smallskip}
\multicolumn{6}{l}{\textbf{C is the most frequent:}}\\
\noalign{\smallskip}\hline\noalign{\smallskip}
Hepatitis C	&	3.8192	&	2894	&	9646	&	2.3334	&	8.7827	\\
Hepatitis D	&	3.1128	&	505	&	1682	&	2.3327	&	13.0101	\\
Hepatitis E	&	3.4866	&	2305	&	7176	&	2.1137	&	8.2778	\\
Phage MS2	&	4.0693	&	934	&	3569	&	2.8223	&	9.9534	\\
\noalign{\smallskip}\hline\noalign{\smallskip}
\multicolumn{6}{l}{\textbf{G is the most frequent:}}\\
\noalign{\smallskip}\hline\noalign{\smallskip}
Yellow fever	&	3.8711	&	3088	&	10862	&	2.5178	&	10.0036	\\
Zika	&	3.8499	&	3140	&	10794	&	2.4379	&	9.3213	\\
\noalign{\smallskip}\hline\noalign{\smallskip}
\multicolumn{6}{l}{\textbf{T is the most frequent:}}\\
\noalign{\smallskip}\hline\noalign{\smallskip}
Feline-CoV	&	3.7072	&	9588	&	29355	&	2.0617	&	6.4845	\\
HCoV-229E	&	3.491	&	9446	&	27317	&	1.892	&	5.9998	\\
HCoV-HKU1	&	3.0233	&	12002	&	29926	&	1.4935	&	3.7750	\\
HCoV-NL63	&	3.0976	&	10806	&	27553	&	1.5499	&	4.0621	\\
HCoV-OC43	&	3.4081	&	10931	&	30741	&	1.8124	&	5.5669	\\
MERS	&	3.7682	&	9800	&	30119	&	2.0735	&	6.4379	\\
SARS	&	3.8944	&	9144	&	29751	&	2.2537	&	7.9013	\\
SARS-CoV-2	&	3.7307	&	9595	&	29903	&	2.1166	&	7.3059	\\
\noalign{\smallskip}\hline
\end{tabular} }
\end{table}
\end{document}